\documentclass[prl,floatfix,preprint,superscriptaddress,showpacs]{revtex4}
\usepackage{graphicx}
\usepackage[english]{babel}
\usepackage[usenames]{color}
\usepackage{amsmath,amsfonts,amssymb,float}
\usepackage{dcolumn}
\usepackage{bm}

\begin{document}

\title{Production of picosecond, kilojoule, petawatt laser pulses via Raman
  amplification of nanosecond pulses}

\author{R.M.G.M. Trines}
\altaffiliation{also at Lancaster University, Lancaster, UK}
\affiliation{Central Laser Facility, STFC Rutherford Appleton Laboratory, Harwell Science and Innovation Campus, Didcot, Oxon, OX11 0QX, United Kingdom}
\author{F. Fi\'uza}
\affiliation{GoLP/Instituto de Plasmas e Fus\~ao Nuclear - Laboratorio
  Associado, Instituto Superior T\'ecnico, 1049-001 Lisbon, Portugal}
\author{R. Bingham}
\altaffiliation{also at University of Strathclyde, Glasgow, UK}
\affiliation{Central Laser Facility, STFC Rutherford Appleton Laboratory, Harwell Science and Innovation Campus, Didcot, Oxon, OX11 0QX, United Kingdom}
\author{R.A. Fonseca}
\altaffiliation{also at DCTI/ISCTE Lisbon University Institute, 1649-026 Lisbon, Portugal}
\author{L.O. Silva}
\affiliation{GoLP/Instituto de Plasmas e Fus\~ao Nuclear - Laboratorio
  Associado, Instituto Superior T\'ecnico, 1049-001 Lisbon, Portugal}
\author{R.A. Cairns}
\affiliation{University of St Andrews, St Andrews, Fife KY16 9AJ,
 United Kingdom}
\author{P.A. Norreys}
\altaffiliation{also at Imperial College London, London, UK}
\affiliation{Central Laser Facility, STFC Rutherford Appleton Laboratory, Harwell Science and Innovation Campus, Didcot, Oxon, OX11 0QX, United Kingdom}
\date\today

\begin{abstract}
  Raman amplification in plasma has been promoted as a means of compressing
  picosecond optical laser pulses to femtosecond duration to explore the
  intensity frontier. Here we show for the first time that it can be used,
  with equal success, to compress laser pulses from nanosecond to picosecond
  duration. Simulations show up to 60\% energy transfer from pump to probe
  pulses, implying that multi-kilojoule ultra-violet petawatt laser pulses can
  be produced using this scheme. This has important consequences for the
  demonstration of fast-ignition inertial confinement fusion.
\end{abstract}

\pacs{52.38.-r, 42.65.Re, 52.38.Bv, 52.38.Hb}
\maketitle

\def\pd#1#2{\frac{\partial #1}{\partial #2}}
\def\pdd#1#2{\frac{\partial^2 #1}{\partial #2 ^2}}
\def\D{\mathrm{d}}
\renewcommand{\vec}[1]{\mathbf{#1}}

The demonstration of fast-ignition (FI) inertial confinement fusion (ICF)
involves two phases: the compression of deuterium-tritium fuel to high density
and the formation of a hot spot region on the side of the fuel at peak
compression via the stopping of energetic (1--3 MeV) electrons generated by an
intense picosecond laser pulse.  \cite{atzeni,honrubia}. Even with the
deployment of different magnetic collimation concepts, between 40 kJ - 100 kJ
of laser energy needs to be delivered within 16 ps to produce an electron beam
with the required properties \cite{atzeni,kemp,norreys,wei,tabak}.
High-energy petawatt beams of 1-10 picosecond duration are difficult to
generate using conventional solid-state laser systems.

Previous studies of Raman amplification have concentrated on reaching the
intensity frontier, which requires ultra-short pulses in the femtosecond
regime \cite{shvets98,shvets99,malkin05,ren07,ping09,trines10,kirkwood11}.
Here we present novel particle-in-cell simulations, supported by analytic
theory, that confirm that Raman amplification of high-energy nanosecond pulses
in plasma can generate efficient petawatt peak power pulses of picosecond
duration with high conversion efficiency (up to 60\%). The scheme can easily
be scaled from $\omega_0$ to 3$\omega_0$ pulses: only the plasma density
needs to be adjusted such that the ratio $\omega_0/\omega_p$ remains fixed
(where $\omega_0 = 2\pi c/\lambda_0$ is the laser frequency, $\lambda_0$ its
wave length, $\omega_p = \sqrt{e^2 n_0/(\varepsilon_0 m_e)}$ is the plasma
frequency, and $n_0$ is the plasma electron density, while $e$,
$\varepsilon_0$ and $m_e$ have their usual meaning). This scheme provides a
new route to explore the full parameter space for the realisation of the fast
ignition inertial confinement fusion concept in the laboratory. This work also
opens up a wide range of other high energy density physics research
applications, including monochromatic K$_\alpha$ x-ray \cite{park}, proton
beam \cite{borghesi} and Compton radiography of dense plasmas
\cite{tommasini}, among many others.


Raman amplification in plasma (characteristic plasma frequency $\omega_p$)
works as follows \cite{shvets98,shvets99}. A long pump laser beam (frequency
$\omega_0$, wave number $k_0$) and a counter-propagating short probe pulse
(frequency $\omega_0-\omega_p$, wave number $\omega_p/c - k_0$) interact via a
longitudinal plasma wave (frequency $\omega_p$, wave number $2k_0 -
\omega_p/c$) via the process known as stimulated Raman backscattering
\cite{forslund}. This causes a large fraction of the energy of the long pump
pulse to be transferred to the short probe pulse. Because the amplified probe
is normally up to 1000 times shorter than the pump, its final power can
be hundreds of times higher than that of the pump beam.

\begin{figure}[ht]
\includegraphics[width=0.45\textwidth]{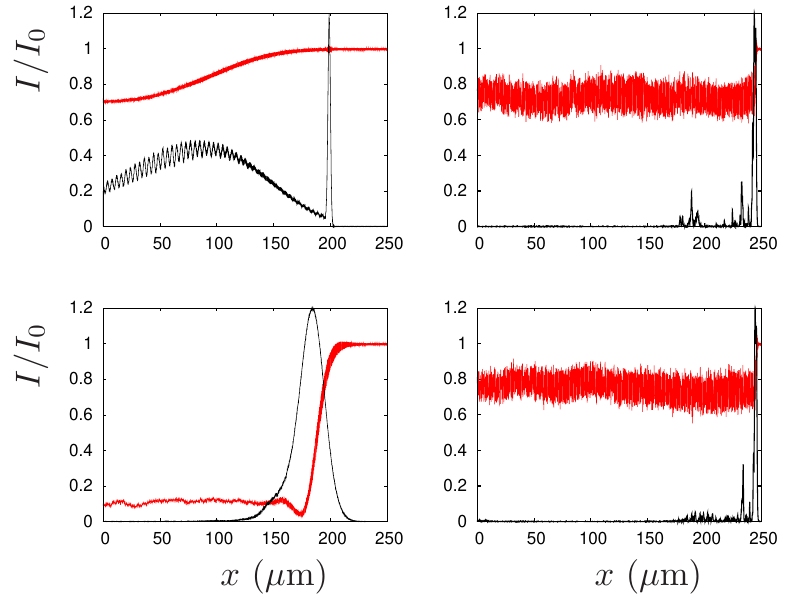}
\caption{Effects of pump intensity of growth of long and short probes. Shown
  are the relative intensities of pump (red) and probe (black) versus
  longitudinal coordinate in $\mu$m. Top row: 50 fs probe and pump with $a_0 =
  0.01$, 0.1 respectively. The less intense pump causes the short probe to
  stretch, while the more intense pump causes the probe to remain short.
  Bottom row: Same as before, but with 500 fs probe. At low pump intensity the
  probe remains long, while at higher pump intensity the probe is shortened.}
\label{fig:1}
\end{figure}

Raman amplification of ultra-short ($\sim 25$ fs) pulses at high intensities
was the subject of an extensive recent investigation \cite{trines10}. In the
case of FI in ICF, however, the compression of long (nanosecond) pump beams to
medium (picosecond) duration is needed. This cannot be done by simply
extending the scheme of Ref. \cite{trines10} (pump intensities of
$10^{14}-10^{15}$ W/cm$^2$) to longer pump pulses because of the increasing
influence of pump and probe instabilities, probe saturation and probe
shortening at longer interaction lengths. However, it follows from the
self-similar theory of Raman amplification developed by Malkin, Shvets and
Fisch \cite{shvets99} that ns-to-ps compression
 can be accomplished by reducing the intensities of pump and probe.
Although the self-similar theory has been used to predict the self-contraction
of the probe \cite{shvets99,kim03,clark03a,clark03b,malkin05}, it has not been
used before to increase the final probe duration. We introduce the
dimensionless pulse amplitude $a \equiv 8.55\times 10^{-10} \sqrt{g} \sqrt{I
  \lambda^2 [\mathrm{Wcm}^{-2}\mu \mathrm{m}^2]}$, where $I$ and $\lambda$
denote the peak intensity and wave length of the laser beam (pump or probe)
under consideration and $g=1$ ($g=1/2$) denotes linear (circular)
polarisation, and write $a_0$ ($a_1$) to denote the amplitude of the pump
(probe) pulse. Following Malkin \emph{et al.}  \cite{shvets99}, we find that
the probe duration after amplification is given by
\begin{equation}
\label{eq:1}
t_\mathrm{probe} = 2g\xi_M^2/(\omega_0 \omega_p a_0^2 t_\mathrm{pump})
\end{equation}
where $\xi_M$ is a constant of the self-similar interaction, $\xi_M \sim 5$
for a 1 ps probe at 351 nm and $10^{13}$ W cm$^{-2}$, and increases by $\sim
1$ when the intensity decreases by an order of magnitude.  Thus, the probe
duration can be increased by keeping the pump intensity low, even for long
pump pulses. If one fixes $\omega_0/\omega_p = 20$ and
$t_\mathrm{pump}/t_\mathrm{probe} = 1000$ as in \cite{trines10}, then the pump
intensity needed to obtain a certain optimal probe duration is given by:
\begin{equation}
\label{eq:2}
I_\mathrm{pump} = 3.9\times 10^{11} / (t_\mathrm{probe}\mathrm{[ps]})^2
\ \mathrm{[W/cm^2]},
\end{equation}
independent of pump laser wave length. As an example, producing a 2 ps probe
from a 2 ns pump requires an intensity of $1.0\times 10^{11}$ W cm$^{-2}$.

To illustrate this, Figure \ref{fig:1} shows the results of four 1-D
particle-in-cell simulations (using the code XOOPIC \cite{oopic}) where the
pump intensity is either high or low, and the initial probe duration is either
long or short. It is found that a low pump intensity leads to a long final
probe, while a high pump intensity leads to a short final probe, independent
of the initial probe duration.

Limits on the applicability of the self-similar theory are posed by the pump
intensity, compression ratio and the frequency ratio $\omega_0/\omega_p$. The
pump intensity should not exceed the value given by (\ref{eq:2}) by much, and
the compression should not be (much) larger than 1000, both to avoid damage by
instabilities. Although increasing $\omega_0/\omega_p$ will increase the
duration of the probe, it will also facilitate wave breaking of the Langmuir
wave that couples pump and probe, disrupting the amplification process
\cite{shvets99}; using $\omega_0/\omega_p \leq 20$ is recommended to prevent
this.

Lowering the pump intensity will increase the duration of the amplified probe
even further. This has been explored in a series of full particle-in-cell
(PIC) numerical simulations using the codes XOOPIC and OSIRIS \cite{osiris}.
In all simulations, the central pump wave length was 351 nm, the plasma
density was $2.3\times 10^{19}$ cm$^{-3}$ ($\omega_0/\omega_p = 20$), and the
plasma was initially cold with static ions. In order to infer the effects of
the ion motion in the amplification of long pulses, simulation (III) below was
repeated with mobile ions (simulation III'). The amplified pulse achieved
similar levels of amplification and a good final pulse shape, justifying the
use of static ions in the other simulations.

\begin{table}[ht]
\begin{tabular}{c||c|c|c|c|c|c|c}
& I & II & III & III' & IV & V & VI \\
\hline
$a_0$ & 0.0044 & 0.003 & 0.0016 & 0.0016 & 0.001 & 0.001 & 0.00056 \\ 
$a_1$ & 0.0044 & 0.003 & 0.0044 & 0.0044 & 0.003 & 0.001 & 0.00056 \\
$t_\mathrm{pu}$ (ps) & 100 & 133 & 100 & 100 & 67 & 133 & 133 \\
$t_\mathrm{pr}$ (fs) & 65 & 28 & 230 & 330 & 400 & 283 & 2180 \\
Eff.(\%) & 20 & 39 & 50 & 40 & 60 & 44 & 7 \\
$\xi_M$ & 14.2 & 7.0 & 8.9 & 10.6 & 6.2 & 7.3 & 11.5 \\
\hline
\end{tabular}
\caption{Summary of our simulation results. In each simulation, $\lambda_0 =
  351$ nm and $\omega_0/\omega_p = 20$ were used. For each simulation, the
  initial pump ($a_0$) and probe ($a_1$) amplitudes, initial pump
  ($t_\mathrm{pu}$) and final probe ($t_\mathrm{pr}$) duration, and energy
  transfer efficiency are given, as well as $\xi_M = (a_0^2 \omega_0 \omega_p
  t_\mathrm{pu} t_\mathrm{pr})^{1/2}$, to verify compliance with the
  self-similar theory. Simulation (III') is similar to (III), only using
  mobile ions.}
\label{table:1}
\end{table}

The summary of our simulation results is shown in Table \ref{table:1}.
Overall, the final probe duration is found to increase with decreasing pump
amplitude, and to decrease with increasing pump duration. For simulations
(II)-(V), the efficiency is 40-60\% and the self-similar parameter $\xi_M$
ranges from 6 to 9, in reasonable agreement with the theoretical prediction $5
< \xi_M < 7$.  The poor efficiency in (I) is caused by the combination of high
pump intensity and long interaction length, triggering premature pump RBS and
probe saturation. The poor efficiency in (VI) is caused by the long start-up
time (see below) resulting from the low pump and probe intensities. These
simulations also exhibit a relatively high value for $\xi_M$, showing that the
probe has not yet fully entered the self-similar regime (VI) or has already
left it (I).

The amplification of the probe may be affected by a number of instabilities:
Raman forward scattering, modulational instability, filamentation, and
parasitic RBS of the pump before it meets the probe. Full-scale
multi-dimensional PIC simulations are needed to investigate such instabilities
properly, see e.g. Ref. \cite{trines10}, but this may not be practical given
the interaction distances involved (e.g. 150 mm for an 1 ns pump).
Nevertheless, the impact of the filamentation, modulational and Raman forward
scattering instabilities on the growing probe can be estimated as follows.
From Ref. \cite{shvets99}, we find that the characteristic growth times for
these instabilities are given by $t_\mathrm{fw} \propto 1/(a_0
\omega_p^{3/2})$ and $t_\mathrm{md} \propto 1/(a_0^{4/3} \omega_p)$. For a
fixed $\omega_0$, $\omega_p$ and compression ratio, e.g. $t_\mathrm{probe} =
t_\mathrm{pump}/1000$, we have $t_\mathrm{probe} t_\mathrm{pump} \propto
t_\mathrm{pump}^2 \propto 1/a_0^2$. Then $t_\mathrm{pump}/t_\mathrm{fw}$ does
not depend on $a_0$, while $t_\mathrm{pump}/t_\mathrm{md} \propto a_0^{1/3}$,
i.e.  this ratio improves for decreasing $a_0$. For the filamentation
instability in the short-pulse limit, Max \emph{et al.} \cite{max74} or
Bingham and Lashmore-Davies \cite{bing76} provide a growth rate of $\gamma =
(a_0^2/8)(\omega_p^2/\omega_0)$, so $t_\mathrm{fil} \propto 1/a_0^2$ and
$t_\mathrm{pump}/t_\mathrm{fil} \propto a_0$, once again improving for
decreasing $a_0$. In short, Raman amplification of long pulses at low
intensities suffers less from damaging instabilities than it does for short
pulses at high intensities.

\begin{figure}[ht]
\includegraphics[width=0.45\textwidth]{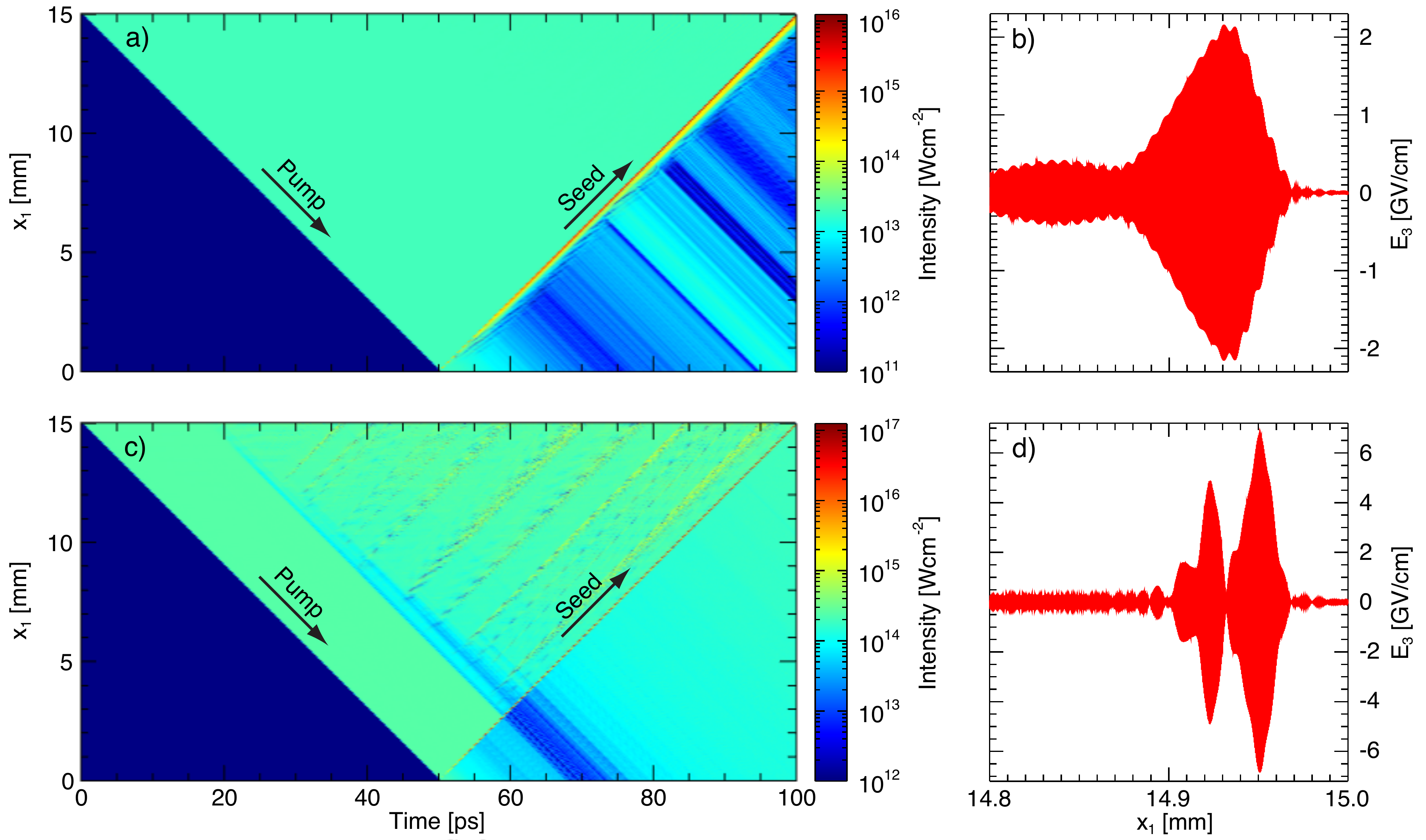}
\caption{Raman amplification of a 100 ps pump to obtain a $\sim 200$ fs probe,
  for a pump wave length of 351 nm. (a) Intensity versus position and time for
  the entire pump-probe interaction, for a pump intensity of $2.7\times
  10^{13}$ W/cm$^2$, obtained using Eq. (\ref{eq:2}). For this intensity, the
  pump propagation is stable with hardly any precursors to the relatively long
  probe, while the pump is efficiently depleted. (b) Transverse electric field
  of the final probe for the intensity under (a). (c) As (a), but for a pump
  intensity of $2\times 10^{14}$ W/cm$^2$. Premature pump RBS generates
  precursors to the probe pulse, which is much shorter while the pump
  depletion is less. (d) Transverse electric field of the final probe for the
  intensity under (c).}
\label{fig:3}
\end{figure}

Simulations of the entire plasma column have been conducted, using OSIRIS, to
study the stability of a 100 ps pump pulse as it traverses a 15 mm plasma
column before it meets the probe. Figure \ref{fig:3} shows the results from
simulations (I) and (III) in Table \ref{table:1}, where (III) has a pump
intensity of $2.7\times 10^{13}$ W/cm$^2$, obtained using Eq. (\ref{eq:2}),
while (I) has a much higher intensity of $2\times 10^{14}$ W/cm$^2$, all other
parameters identical. For the lower intensity, both pump propagation and probe
amplification are stable, and a smooth probe is obtained without precursors.
For the higher intensity, the pump suffers from parasitic instabilities,
mostly premature RBS, leading to probe precursors, and the probe is shorter
while its envelope is less smooth, which is partly caused by the probe taking
on the characteristic multi-period $\pi$-pulse shape. This emphasises the
importance of keeping the intensity at or below the value predicted by Eq.
(\ref{eq:2}), when the Raman amplification of a long pump is required.

One issue that is not immediately obvious from the self-similar theory is the
existence of a ``start-up period'' for Raman amplification of low-intensity
pulses. The simulations show that when both pump and probe amplitudes are
below $a_0 = 0.01$ ($\sim 10^{14}$ W/cm$^2$ for a $\sim 1\ \mu$m pump wave
length), the pump and probe need to interact over several mm before the probe
amplification starts in earnest. For a 2 ps initial probe duration, this
ranges from less than 2 mm for $a_0 = a_1 = 0.001$ and 10 mm for $a_0 = a_1 =
0.000562$ to 20 mm for $a_0 = a_1 = 0.0003$. This follows from the fact that
the initial probe amplitude and duration for the ideal self-similar solution
are related as $t_\mathrm{probe} \approx 4\times 10^{-15}/a_1$ ($\lambda_0= 351$
nm, $\omega_0/\omega_p = 20$). Thus, a 2 ps initial probe duration ideally
requires $a_1 = 0.002$ initially. In our simulations, however, $a_1$ is well
below that value and the actual probe requires an increasing amount of time to
evolve into a self-similar probe for decreasing $a_1$. Hence the start-up
period, whose length also depends on the pump intensity. This effect can be
mitigated by increasing the initial probe intensity and/or duration, thus
ensuring that the start-up period remains a limited fraction of the pump
duration and the overall efficiency remains reasonable even for very low pump
intensities.
\begin{figure}[ht]
\includegraphics[width=0.45\textwidth]{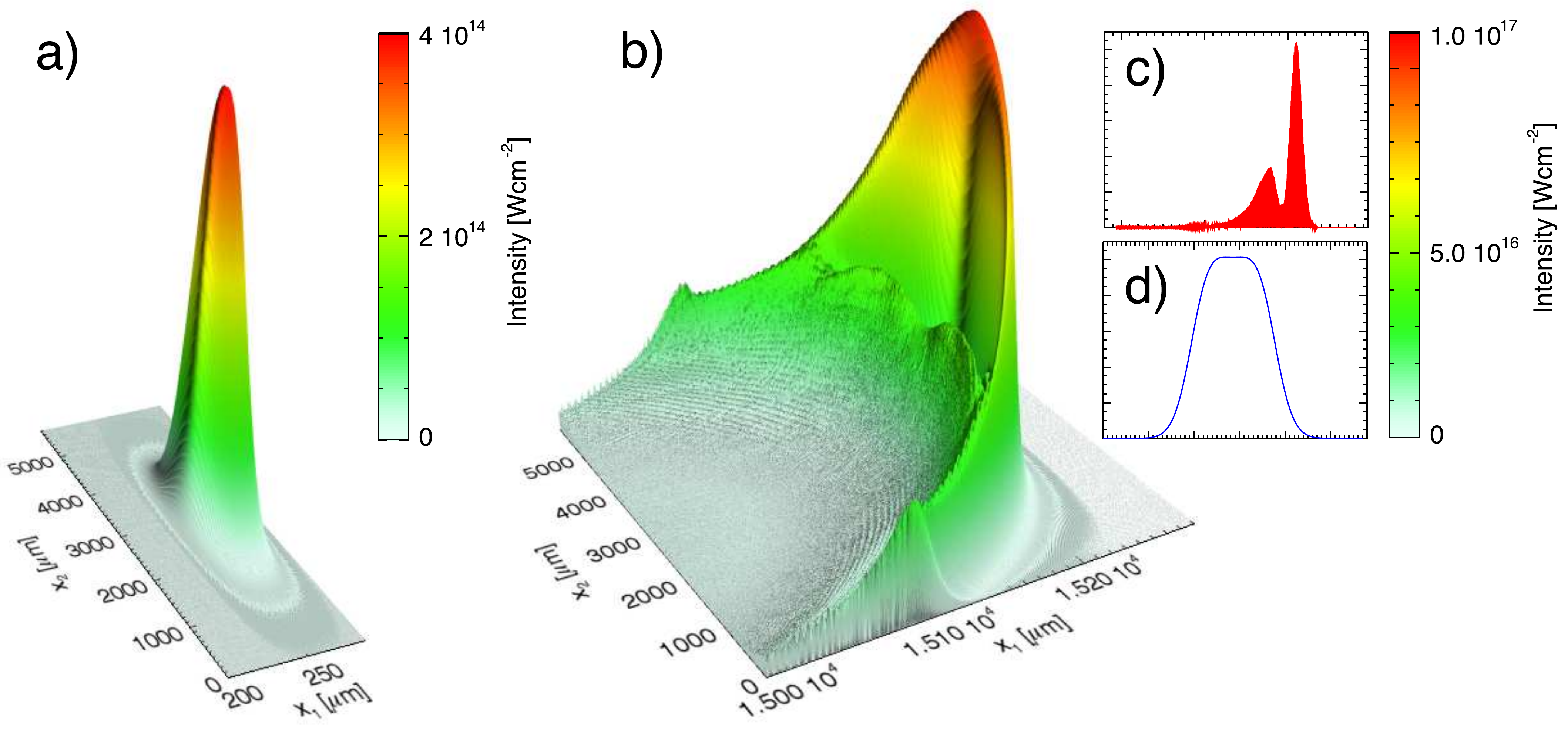}
\caption{Raman amplification of a 100 ps long pump to obtain a $\sim 150$ fs,
  mm wide, 2 PW probe, for a wave length of 351 nm. (a) Initial and (b) final
  intensity profile of the probe pulse after amplification in a 1.5 cm plasma
  column. Insets show central lineouts of the amplified pulse in the (c)
  longitudinal and (d) transverse directions.}
\label{fig:4}
\end{figure}

In order to study the stability of the process in multiple dimensions, where
transverse instabilities can come into play, we have performed a 2D OSIRIS
simulation using the parameters of simulation (III) and a probe spot FWHM of
1.2 mm. This relatively wide probe pulse is efficiently amplified to a peak
intensity of $10^{17}$ Wcm$^{-2}$, corresponding to a final power of $\sim
1.5$ PW (Fig. \ref{fig:4}). As predicted for these optimized parameters,
transverse instabilities are controlled and the amplified pulse retains a
smooth envelope. The bowed shape of the amplified pulse is similar to the
shape observed for ultrashort pulses \cite{trines10}, and does not influence
the amplification process. While the final pulse power significantly exceeds
the critical power for self-focusing, $P_c = 17 (\omega_0/\omega_p)^2$ GW =
6.8 TW \cite{sprangle87}, no significant self-focusing occurs during the
interaction, which can be explained as follows. The spot radius $R$ of a
self-focusing laser pulse in plasma is described by $R = R_0
\sqrt{1-z^2/z_0^2}$, where $R_0$ is the minimum spot radius in vacuum and $z_0
= z_R/\sqrt{P/P_c -1}$ is the typical distance for self-focusing, with $z_R =
\pi R_0^2/\lambda$ being the Rayleigh length and $P$ the laser power. For a
1.5 PW laser pulse with $R_0 = 1$ mm and $\lambda = 351$ nm,
$z_0 =  63$ cm, significantly exceeding the typical amplification distances
involved (1.5 -- 2 cm). Higher powers require larger spot sizes, since the
intensity $I$ is fixed by the probe duration and compression ratio; this leads
to $z_R \propto P/I \propto P$ and $z_0 \propto \sqrt{P}$. Thus, the influence
of self-focusing decreases for increasing probe power.

In the above 2-D simulation, a $2.7\times 10^{13}$ W/cm$^2$, 100 ps pump has
been used to obtain a $\sim 150$ fs probe with a smooth envelope. In
simulation (IV), a $\sim 10^{13}$ W/cm$^2$, 67 ps pump has been used to obtain
a 0.4 ps probe.  Furthermore, it was shown in ref. \cite{trines10} that a 25
ps pump at $10^{15}$ W/cm$^2$ (800 nm wave length) can be compressed to about
25 fs while probe modulation and RFS remain at an acceptable level.
Extrapolating these cases via the self-similar theory, we predict that a 2 ps
probe can be obtained using a 2 ns pump at an intensity of $(0.7-1.5)\times
10^{11}$ W/cm$^2$, in good agreement with the value of $1.0\times 10^{11}$
W/cm$^2$ predicted by Eq. (\ref{eq:2}). Assuming the pump beam contains 10 kJ
in 2 ns, the pump power will be 5 TW, so a 50 cm$^2$ cross section is needed
for the interaction.  Such energetic pump beams can be obtained at e.g. the
National Ignition Facility \cite{moses}, the Omega EP laser system at the
Laboratory for Laser Energetics in Rochester \cite{boehly,waxer}, and the
Laser M\'egajoule project \cite{fleurot}.

In summary, we have investigated the Raman amplification and compression of
nanosecond laser pulses to picosecond duration, exploiting the self-similar
properties of the process. We have shown that, for a constant pump-to-probe
compression ratio, the optimal pump and probe durations will increase for
decreasing pump intensity. In addition, we have shown that the relative
importance of undesirable instabilities remains the same (pump RBS, probe RFS)
or even decreases (modulational and filamentation instabilities) with
decreasing pump intensity. Energy transfer efficiencies of up to 60\% have
been found. Thus, Raman amplification in plasma can be used to generate
picosecond pulses of moderate intensity but large total energy.  This has
important consequences for a wide range of applications in high energy density
physics, particularly fast-ignition ICF and x-ray and proton radiographic
diagnosis of dense plasmas. Most importantly, our approach provides a
potential route to the full-scale demonstration of fast ignition inertial
confinement fusion using existing facilities.

Work supported by STFC's Central Laser Facility and Centre for Fundamental
Physics, by EPSRC through grant EP/G04239X/1, by the European Research Council
(ERC-2010-AdG Grant 267841), and by FCT (Portugal) grants PTDC/FIS/111720/2009
and SFRH/BD/38952/2007. We thank C. Joshi, N. Fisch and R. Kirkwood for
stimulating discussions, N. Loureiro for the moving-window antenna in OSIRIS,
UC Berkeley for the use of XOOPIC and the OSIRIS consortium for the use of
OSIRIS. We acknowledge the assistance of HPC resources (Tier-0) provided by
PRACE on Jugene (Germany). Simulations were performed on the Scarf-Lexicon
Cluster (STFC RAL), the IST Cluster (IST Lisbon), the Hoffman cluster (UCLA)
and the Jugene supercomputer (Germany).


\begin{thebibliography}{99}
\bibitem{atzeni} S. Atzeni \emph{et al.}, Phys. Plasmas {\bf 15}, 056311
  (2008).
\bibitem{honrubia} J. Honrubia and J. Meyer-ter-Vehn, Plasma
  Phys. Control. Fusion {\bf 51} 014008 (2009).
\bibitem{kemp} A.J. Kemp, Y. Sentoku and M. Tabak, Phys. Rev. E {\bf 79},
  066406 (2009).
\bibitem{norreys} P.A. Norreys \emph{et al.}, Nucl. Fusion {\bf 49}, 104023
  (2009).
\bibitem{wei} M.S. Wei \emph{et al.}, Phys. Plasmas {\bf 15}, 083101 (2008).
\bibitem{tabak} M. Tabak \emph{et al.}, Phys. Plasmas, {\bf 1}, 1626 (1994).
\bibitem{shvets98} G. Shvets \emph{et al.}, Phys. Rev. Lett. {\bf 81},
  4879-4882 (1998).
\bibitem{shvets99} V.M. Malkin \emph{et al.}, Phys. Rev. Lett. {\bf 82},
  4448-4451 (1999).
\bibitem{malkin05} V.M. Malkin and N.J. Fisch, Phys. Plasmas {\bf 12}, 044507
  (2005).
\bibitem{ren07} J. Ren \emph{et al.}, Nature Physics {\bf 3}, 732-736 (2007).
\bibitem{ping09} Y. Ping \emph{et al.}, Phys. Plasmas {\bf 16}, 123113 (2009).
\bibitem{trines10} R.M.G.M. Trines \emph{et al.}, Nature Physics {\bf 7}, 87
  (2011).
\bibitem{kirkwood11} R.K. Kirkwood \emph{et al.}, Phys. Plasmas {\bf 18},
  056311 (2011).
\bibitem{park} H.-S. Park \emph{et al.}, Phys. Plasmas {\bf 13}, 056309 (2006).
\bibitem{borghesi} M. Borghesi \emph{et al.}, Phys. Plasmas {\bf 9}, 2214
  (2002).
\bibitem{tommasini} R. Tommasini \emph{et al.}, Rev. Sci. Instrum. {\bf 79},
  10E901 (2008).
\bibitem{forslund} D.W. Forslund, J.M. Kindel and E.L. Lindman, Phys. Fluids
  {\bf 18}, 1002-1016 (1975).
\bibitem{kim03} J. Kim, H.J. Lee, H. Suk and I.S. Ko, Phys. Lett. A {\bf 314},
  464 (2003).
\bibitem{clark03a} D.S. Clark and N.J. Fisch, Phys. Plasmas {\bf 10}, 4837
  (2003).
\bibitem{clark03b} D.S. Clark and N.J. Fisch, Phys. Plasmas {\bf 10}, 4848
  (2003).
\bibitem{oopic} J.P. Verboncoeur, A.B. Langdon and N.T. Gladd,
  Comp. Phys. Comm. {\bf 87}, 199-211 (1995).
\bibitem{osiris} R.A. Fonseca, L.O. Silva, R.G. Hemker, \emph{et al.,}
  Lect. Not. Comp. Sci. \textbf{2331}, 342-351 (2002).
\bibitem{max74} C.E. Max, J. Arons and A.B. Langdon, Phys. Rev. Lett. {\bf
    33}, 209-212 (1974).
\bibitem{bing76} R. Bingham and C.N. Lashmore-Davies, Nuclear fusion {\bf 16},
  67 (1976).
\bibitem{sprangle87} P. Sprangle, C.-M. Tang, and E. Esarey, IEEE
  Trans. Plas. Sci. {\bf 15}, 145-153 (1987).
\bibitem{moses} E.I. Moses, J. Phys. Conf. Ser. {\bf 112}, 012003 (2008).
\bibitem{boehly} T.R. Boehly \emph{et al.}, Opt. Commun. {\bf 133}, 495 (1997).
\bibitem{waxer} L.J. Waxer \emph{et al.}, Opt. Photon. News {\bf 16}, 30
  (2005).
\bibitem{fleurot} N. Fleurot, C. Cavailler and J.L. Bourgade, Fusion
  Engineering and Design {\bf 74}, 147 (2005).
\end{thebibliography}
\end{document}